\documentclass[aps,prl,twocolumn,superscriptaddress,showkeys]{revtex4-1}

\usepackage{color}

\usepackage{graphicx}
\usepackage{dcolumn}
\usepackage{bm}

\usepackage[english]{babel}
\usepackage[latin1]{inputenc}
\usepackage[bottom=2cm,top=2cm, left=1.5cm, right=1.5cm]{geometry}

\usepackage[unicode=true, bookmarks=true, bookmarksnumbered=false, bookmarksopen=false, breaklinks=false, pdfborder={0 0 1}, backref=false, colorlinks=false]{hyperref}
\hypersetup{pdftitle={Photoluminescence of focused ion beam implanted Er-doped YSO crystals}, pdfauthor={N. Kukharchyk, S. Pal, J. R\"{o}diger, A. Ludwig, S. Probst, A. V. Ustinov, P. Bushev, A. D. Wieck}, pdfkeywords={rare earth, concentration dependance, annealing, activation}}
\usepackage[sort&compress]{natbib}

\begin{document}

\title{Photoluminescence of focused ion beam implanted Er$^{3+}$:Y$_{2}$SiO$_{5}$ crystals}

\author{N. Kukharchyk}
\affiliation{Angewandte Festk\"{o}rperphysik, Ruhr-Universit\"{a}t Bochum, D-44780 Bochum, Germany}
\email{nadezhda.kukharchyk@ruhr-uni-bochum.de}

\author{S. Pal}
\affiliation{Angewandte Festk\"{o}rperphysik, Ruhr-Universit\"{a}t Bochum, D-44780 Bochum, Germany}

\author{J. R\"{o}diger}
\affiliation{RUBION, Ruhr-Universit\"{a}t Bochum, D-44780 Bochum, Germany}

\author{A. Ludwig}
\affiliation{Angewandte Festk\"{o}rperphysik, Ruhr-Universit\"{a}t Bochum, D-44780 Bochum, Germany}

\author{S. Probst}
\affiliation{Physikalisches Institut, Karlsruhe Institute of Technology, D-76128 Karlsruhe, Germany}

\author{A. V. Ustinov}
\affiliation{Physikalisches Institut, Karlsruhe Institute of Technology, D-76128 Karlsruhe, Germany}

\author{P. Bushev}
\affiliation{Experimentalphysik, Universit\"{a}t des Saarlandes, D-66123 Saarbr\"{u}cken, Germany}

\author{A. D. Wieck}
\affiliation{Angewandte Festk\"{o}rperphysik, Ruhr-Universit\"{a}t Bochum, D-44780 Bochum, Germany}

\date{\today}

\begin{abstract}
Erbium doped low symmetry Y$_2$SiO$_5$ crystals attract a lot of attention in perspective of quantum information applications. However, only doping of the samples during growth is available up to now, which yields a quite homogeneous doping density. In the present work, we deposit Er$^{3+}$-ions by the focused ion beam technique at Yttrium sites with several fluences in one sample. With a photoluminescence study of these locally doped Er$^{3+}$:Y$_2$SiO$_5$ crystals, we are able to evaluate the efficiency of the implantation process and develop it for the highest efficiency possible. We observe the dependence of the ion activation after the post-implantation annealing on the fluence value.

\end{abstract}

\maketitle

Rare-earth (RE) doped materials are in the focus of interest for modern physics due to their specific properties, such as the presence of high coherence transitions inside the 4f shell and long optical and microwave coherence times. The rapid progress with the REs has been expressed in number of dramatic achievements: the demonstration of long living optical holes in europium-doped yttrium silicate~\cite{Schiller2011}, optical quantum memory based on Pr$^{3+}$:Y$_2$SiO$_5$ crystal for visible wavelengths~\cite{Sellars2010}, storage and generation of entanglement between neodymium doped crystals~\cite{Clausen2011, Usmani2012}. 

Among other REs, the erbium doped crystals are outstanding due to the presence of optical transitions inside the telecom C-band at the wavelength of around 1530-1565 nm. Moreover, the Er$^{3+}$:Y$_2$SiO$_5$ (Er:YSO) crystal possesses the longest measured optical coherence time of about 6 ms among all other solid state systems~\cite{Bottger2002, Bottger2003}. On the hand of the magnetic properties, erbium ions possess a very large magnetic moment of $7\mu_B$. Electron Spin Resonance (ESR) studies revealed that the T$_2$ coherence time can reach the huge value of 100~ms~\cite{Hastings2008}. The development of high fidelity and long-life quantum  memory will boost the implementation of long distance quantum communication protocols in already existing fiber optical networks~\cite{Probst_PRL13, Bushev2011}. 

\begin{figure}[ht!]
\includegraphics[width=0.9\columnwidth]{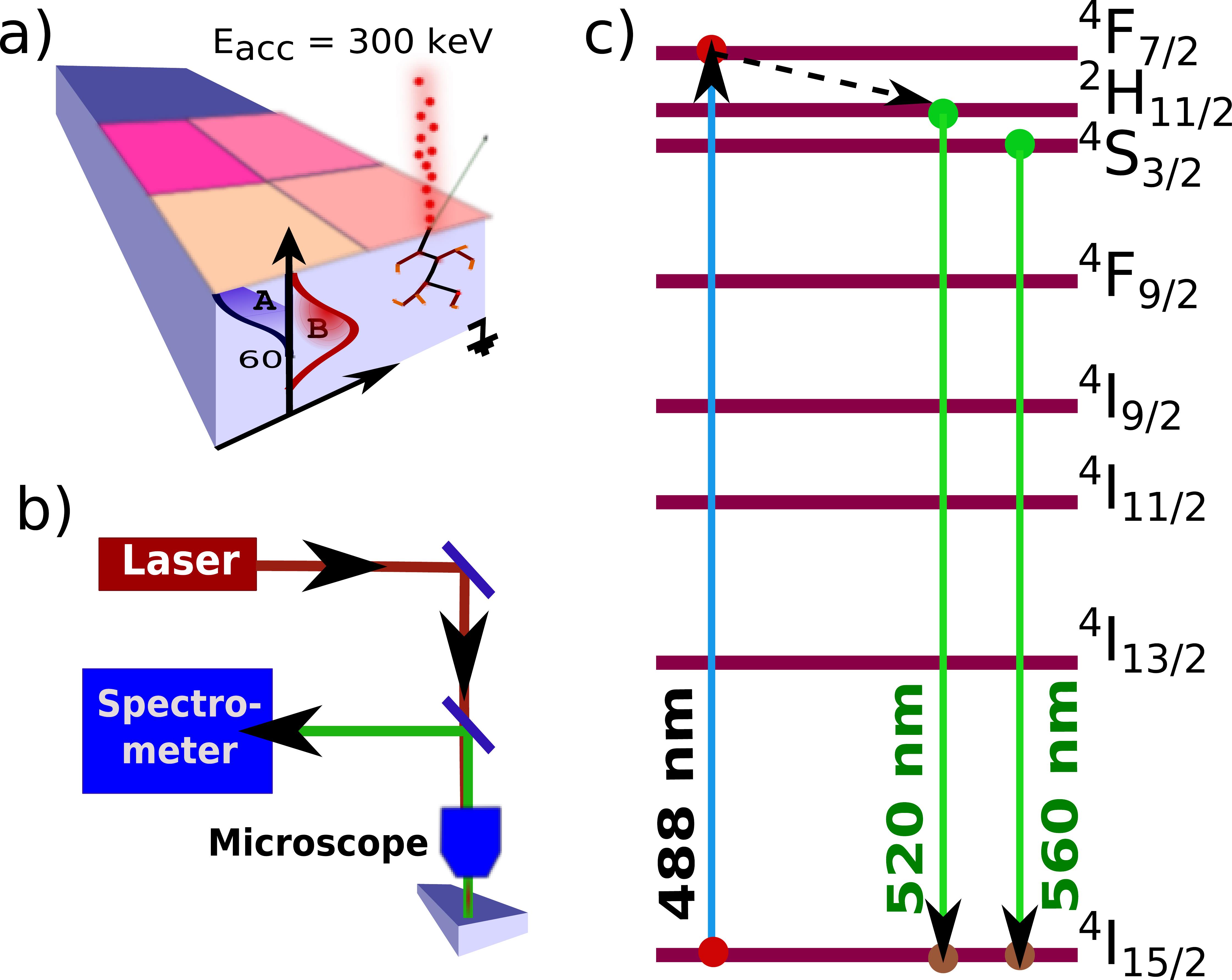}
\caption{\textbf{(a)} Schematic illustration of the implantation process: ion beam generates a collision cascade in the matrix. Curve A corresponds to the damage distribution in the substrate, curve B depicts predicted distribution of the implanted ions. The maximum of the ion distribution peak (for the energy of $300~keV$) is located 60 nm from the surface. \textbf{(b)} Schematics for the confocal setup. The luminescence is measured in reflection. \textbf{(c)} Diagram of the optical multiplets of $Er^{3+}$ ion. Excitation is performed from the ground state $^4I_{15/2}$ to the excited state $^4F_{7/2}$ with 488 nm laser irradiation. Luminescence transitions were detected from the lower neighboring states $^2H_{11/2}$ and $^4S_{3/2}$ to the ground state.}
\label{fig:Setup}
\end{figure}

So far, all of these research deals with the doped-as-grown Er$^{3+}$:Y$_2$SiO$_5$ crystals. However, implementation of Focused Ion Beam (FIB) technique for doping of the YSO crystals looks as a rather handy possibility to perform different types of quantum elements and quantum gates on one chip without mask- and alignment processes. It is especially advantageous in circuit Quantum Electrodynamics (QED) implementations, where one wants to separate the memory elements from the computation elements on the chip. FIB possesses a wide range of flexible parameters, which allow to perform implantation with controlled fluences at different depths at selected positions. Proper adjustment of these implantation parameters, as well as annealing procedure, makes it possible to implant the Erbium ions in the Yttrium sites and recover the distorted site-symmetry. In this letter, we discuss the results of FIB implantation of the Y$_2$SiO$_5$ crystals with the Erbium ions.  

The implantation was performed into a nominally undoped YSO crystal, supplied by the Scientific Materials Inc. Doped as-grown Er:YSO 0.005~\% crystal was taken as reference. All the crystals have the same orientation of optical axes. Purity of the undoped sample is 99.999~\%, which corresponds to the highest possible in-situ presence of 0.0001~\% of Erbium ions. 

The nominally undoped YSO crystal was implanted in an EIKO FIB system with Erbium ions of 300 keV of energy. Ions were extracted from a Liquid Metal Ion Source (LMIS), developed by A. Melnikov et. al. \cite{melnikov2002}. The LMIS alloy consisted of erbium, gold and silicon with $10 \%$ of Erbium. Ion separation was done with a built-in Wien mass-filter, and a resolution of 2~u. There are 6 stable isotopes of Erbium in total, three of them being most abundant: $^{166}$Er $\approx 33\%$, $^{167}$Er $\approx 23\%$ and $^{168}$Er $\approx 27\%$. $^{167}$Er is the only isotope with the nuclear spin of $I = 7/2$, which would result in additional hyperfine splitting of the electron energy levels. This splitting is preferred to be avoided. Though the resolution did not allow to obtain 100\% pure isotopes, we were able to choose position on the interlocked peak with the lowest content of the  $^{167}$Er isotope and the highest content of the $^{166}$Er. The FIB's Pattern Drawing System (PDS) gives a picture resolution of 32 nm with a maximum single pattern area of 1 mm by 1 mm. Four areas with different fluences were performed on one surface, as shown in Fig.~\ref{fig:Setup}~(a).

Figure~\ref{fig:Setup}~(a) schematically shows the collision cascade in the substrate and an approximate distribution of the induced damage (A) and the implanted ions (B). Ion distribution maximum, simulated with SRIM software \cite{Ziegler2010}, is in a depth of 60 nm with a straggle of 39 nm. As the collision interactions between the incident ions and the lattice ions generate a lot of defects, the annealing serves as a necessary tool to repair the defects and restore the crystal structure. We found that the common Rapid Thermal Annealing (RTA) is not efficient enough for the implanted YSO crystals. Therefore, we performed longer thermal annealing procedure. The sample was annealed for 2 hours at 1200$^\circ$C.

The YSO crystal was implanted with four fluences, the values given in the table~\ref{table:tab1}. The volume concentration for each area is estimated for distribution of implanted ions in a 60 nm thick layer. The non-implanted area of the sample was taken as background to eliminate any possible input from in-situ impurities. Doped-as-grown sample was taken as reference to compare with other experimental data.

\begin{table}[ht]
\caption{Ion numbers and fluences} 
\centering      
\begin{tabular}{c || c | c | c |}  
\hline\hline                     
 Sample & Fluence, &  Volume concentration,  & Number of ions*  \\ [0.3ex]
& cm$^{-2}$ & cm$^{-3}$ &\\[0.5ex]
\hline
 & & & \\                   
YSO4**&	4.06$\times10^{13}$	&6.77$\times10^{18}$	&101550\\
	&1.02$\times10^{14}$	& 1.69$\times10^{19}$	&  253875\\
	&2.44$\times10^{14}$	& 4.06$\times10^{19}$	&  609300\\
	&4.87$\times10^{14}$ &	8.12$\times10^{19}$	&  1218600\\ [1ex]
grown	& &	4.78$\times10^{17}$	&  119609\\ [1ex]
undoped	& &	9.57$\times10^{15}$	&  2392\\
\hline 
\end{tabular} \\
*per spot $500 nm \times 500 nm \times 1 \mu m $ \\
**``4'' refers to nomenclature of the sample in this work
\label{table:tab1} 
\end{table}

The sample was characterized optically with a confocal photo-luminescence setup equipped with an Ar-Laser with a wavelength of 488 nm (Fig.~\ref{fig:Setup}~(b)). The confocal spot had a diameter of about 350 nm, with an axial resolution of 1.5 um. The detection range was from 450 nm to 900 nm with a spectral resolution of 0.22 nm. The sample was measured before and after annealing. The wavelength of 488 nm excites the direct transition to the $^4F_{7/2}$ level (Fig.~\ref{fig:Setup}~(c)). Luminescent transitions are observed from the multiplets $^4S_{3/2}$ and $^2H_{11/2}$ to the ground multiplet $^4I_{15/2}$.

\begin{figure}[ht!]
\includegraphics[width=1\columnwidth]{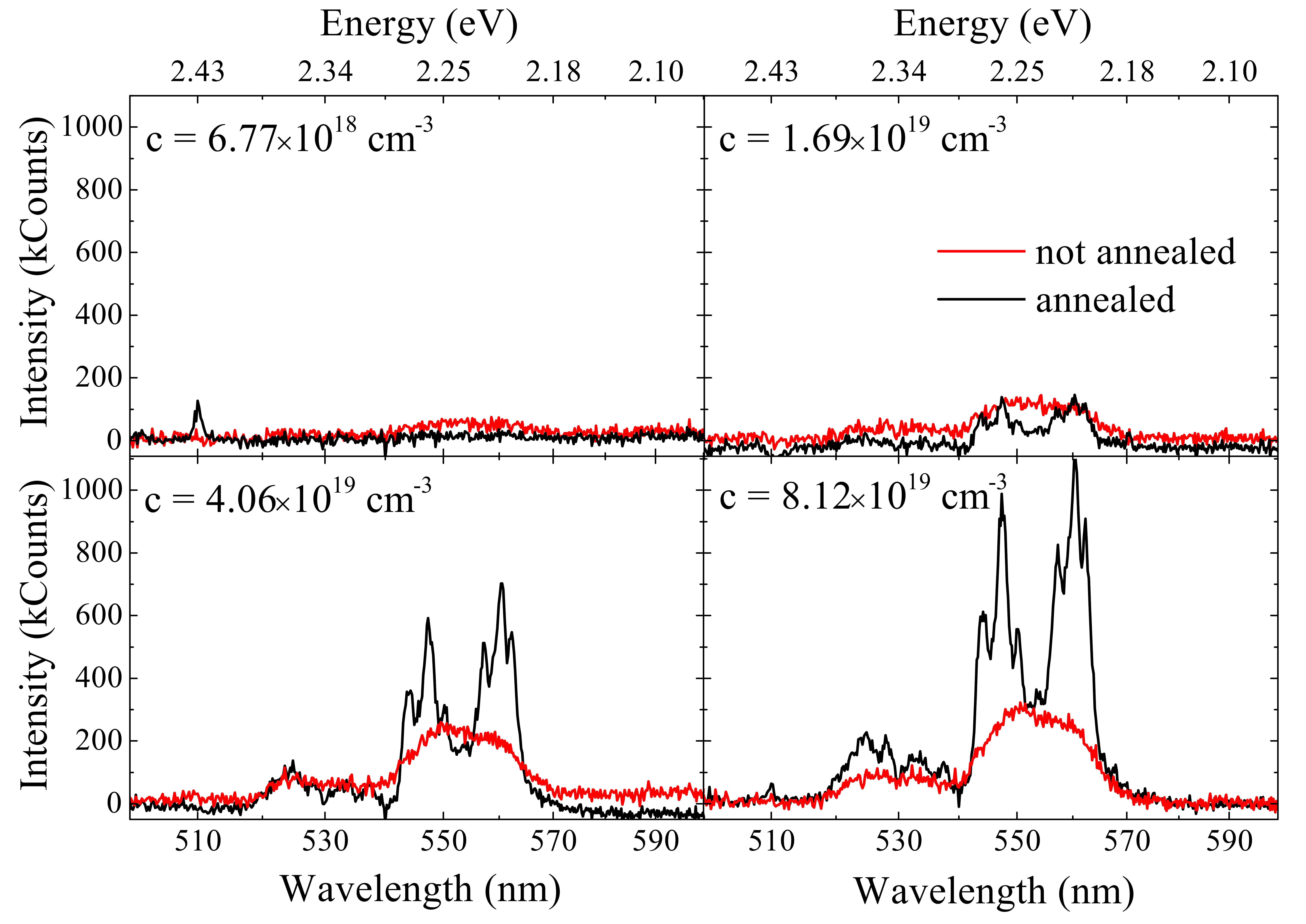}
\caption{(Color online) Comparison of the luminescence of the implanted areas before and after annealing. The registered intensity is dependent on the concentration of the ions in the substrate. Table~\ref{table:tab2} gives a ratio comparison of the intensities before and after the sample was annealed.}
\label{fig:notan}
\end{figure}

Though the f-electrons are ``shielded'' by the outer shells from the surrounding environments, they are still influenced by the symmetry of the crystal field. As the result, each multiplet splits into a number of sub-levels~\cite{Doualant1995, Li1992}. The energy of each sub-level and the probability ($\sim$~intensity) of the luminescent transition to the ground state is strongly dependent on the symmetry of the crystal field~\cite{Ohare1977}. 

In homogeneous - equal symmetry - crystal surroundings, the line-shapes of the optical transitions appear to be also identical. This effect is clearly seen from the comparison of the spectra taken before and after annealing. In Fig.~\ref{fig:notan}, spectra of four implanted areas are given in comparison. The red line corresponds to the sample before annealing. Even before appropriate activation is performed, the implanted ions are capable of giving luminescent response. The line-shape comes in the form of a rather broad transition. When the same areas are measured after annealing (Fig.~\ref{fig:notan}, black line), the manifolds gain structured shape with strongly pronounced independent lines. This effect is a direct indication of the crucial role of the symmetry surrounding. In the non-annealed sample, ions meet a range of different field symmetries induced by the damaged crystal lattice. Alternately, differences in the field symmetries induce differences in the fine structure of the multiplets. As the result, mixture of various manifolds is observed as a broad transition. Annealing repairs the distorted crystal structure and brings ions back to the lattice positions, embedding erbium ions in the yttrium sites. As the symmetry surrounding for the implanted ions becomes homogeneous, repeated distinguished transitions are observed.

\begin{figure}[ht!]
\includegraphics[width=1\columnwidth]{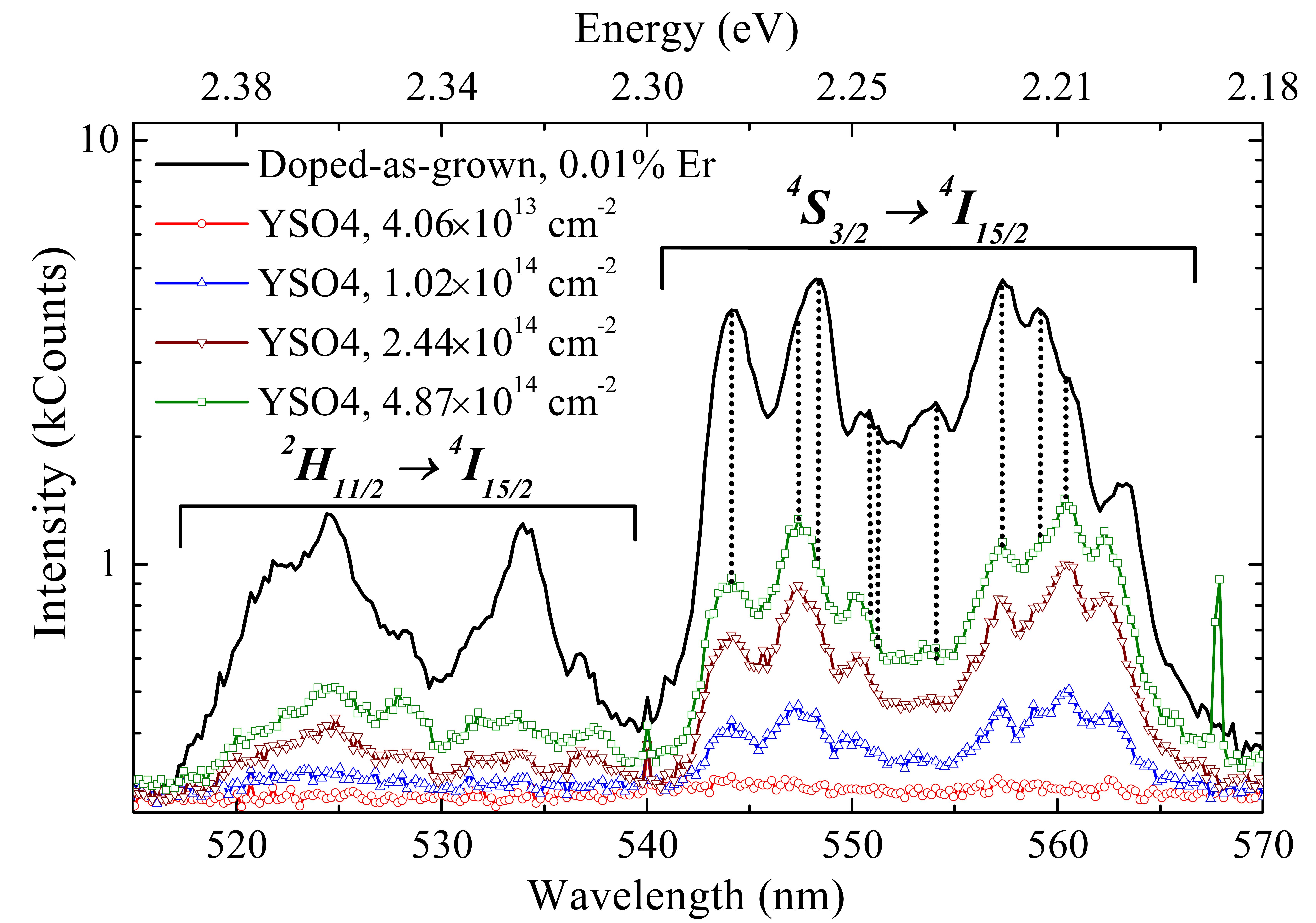}
\caption{(Color online) Spectra of two manifolds  $^4S_{3/2} \rightarrow ^4I_{15/2}$ and $^2H_{11/2} \rightarrow ^4I_{15/2}$ of the implanted sample in comparison to the grown sample. Dotted lines show how the peaks of the implanted sample are related with the transitions in the doped as-grown sample at the same wavelength.}
\label{fig:spec}
\end{figure}

In our study, we compare the shapes of the manifolds $^4S_{3/2} \rightarrow ^4I_{15/2}$ and $^2H_{11/2} \rightarrow ^4I_{15/2}$ of the implanted sample to the doped as-grown. 

In Fig.~\ref{fig:spec}, spectra of the annealed sample are compared to the doped as-grown one. Temperature broadening of the lines does not allow us to distinguish and associate the sub-transitions. However, we can compare the strongest peaks to conclude the symmetry equivalence. As it is easily seen from Fig.~\ref{fig:spec}, symmetry has recovered rather well with appropriate energies for the distinguishable transitions. Still, certain mismatches in the intensities of the sub-transitions indicate the presence of non-repaired defects in the crystal. 

We integrated the intensities over both manifolds $^4S_{3/2} \rightarrow ^4I_{15/2}$ and $^2H_{11/2} \rightarrow ^4I_{15/2}$ together in the wavelength-range of $510~nm - 570~nm$. The relation between the integral intensities for the same area before and after the sample was annealed shows how effectively ions are imbedded in the crystal sites - annealing activation.
Table~\ref{table:tab2} represents the ratio for each of the fluences of the YSO4. It appears that for the fluences higher than $2 \times 10^{14}~cm^{-2}$, the integral intensity increases by a factor of 2.00 - 2.10 after annealing. Whereas for the fluences lower than $2 \times 10^{14}~cm^{-2}$, the luminescence strongly decreases with the decrease of the fluence value. We see this alteration in the value of the ratio to be in direct connection to the implantation process itself: damage in the crystal. Yttrium ions themselves are heavy and cannot be easily substituted by the Erbium ions. From a certain value of the fluence, a stable probability of substitution of yttrium ions per implanted erbium ion is reached. If this value is not reached, a much smaller amount of erbium ions is probable to be embedded in the required site. Therefore, for the higher fluences we observe a stable ratio of the annealed to the non-annealed intensities - as enough interactions take place to replace the yttrium ions with a stable statistics. 

\begin{table}[ht]
\caption{Annealed/non-annealed intensities ratios} 
\centering      
\vspace{1mm}
\begin{tabular}{c || c | c | c | c }  
 Fluence & 4.06$\cdot 10^{13}$ & 1.02$\cdot 10^{14}$ & 2.44$\cdot 10^{14}$ & 4.87$\cdot 10^{14}$ \\ 
\hline        

Ratio    & 0.54 & 0.86 & 2.03 & 2.09	\\

\end{tabular}
\label{table:tab2}  
\end{table}

In order to obtain a more precise rate of  the activated ions, we fit the obtained intensities as a function of the volume concentration, which is typically described~\cite{kalinkin2002} as: 
\[ I = I_s \varphi_k (1 - 10^{- k l c}), \]
where $I_s$ is intensity of the incident irradiation, $\varphi_k$ is the quantum yield coefficient, $k$ is the absorption coefficient of the material, $l$ is the thickness of the excited layer and $c$ is the concentration of the impurity. For $klc < 0.005$, the intensity depends linearly on the concentration. However, if the number of the luminescent impurities increases, non-radiating transitions and interactions may take place, which yields concentration damping, power saturation, shielding effect etc. As the result, the dependency becomes nonlinear. Concentration damping takes place due to the ion-ion interactions within the implanted ions. This can be considered via the quantum yield coefficient~\cite{kalinkin2002}:
\[ \varphi = \varphi_0 \exp(- b (c - c_0)),\]
where $c_0$ is a threshold concentration for which the damping becomes relevant. However, in the range of concentrations performed, we see that the concentration damping is a less probable reason of the nonlinear dependence. Mainly, a non-linear behavior appears from a nonlinear activation rate of annealing of different fluences. As it also appeared in fitting the function, the concentration damping term can be neglected in this range of data. The experimental data were thus fitted with the function \[ I = k_1 (1 - 10^{- k_2 (c-c_0)}).\]

\begin{table}[ht]
\caption{Parameters for fitting function for intensity-on-concentration dependencies} 
\centering      
\vspace{1mm}
\begin{tabular}{ c || c | c | c }  
 Parameter & Not annealed & \parbox[c]{2cm}{Annealed, experimental} & \parbox[c]{2cm}{Annealed, corrected} \\ 
\hline \hline        
k$_1$, kCounts & 49.05 & 112.20 & 110.67	\\
c$_0$, cm$^{-3}$ & 7884.63 & 6.57$\times 10^{18}$ & 8987.54 \\
k$_2$, cm$^3$ & 5.77$\times 10^{-21}$ & 5.34$\times 10^{-21}$ & 4.95$\times 10^{-21}$ \\
\hline
\parbox[c]{2cm}{Concentration, for which  $k_1 x = 0.05$} & 8.67$\times 10^{18}$ & 9.36$\times 10^{18}$ & 1.01$\times 10^{19}$ \\
\hline \hline
\end{tabular}
\label{table:tab3}  
\end{table}

\begin{figure}[htb!]
\includegraphics[width=1\columnwidth]{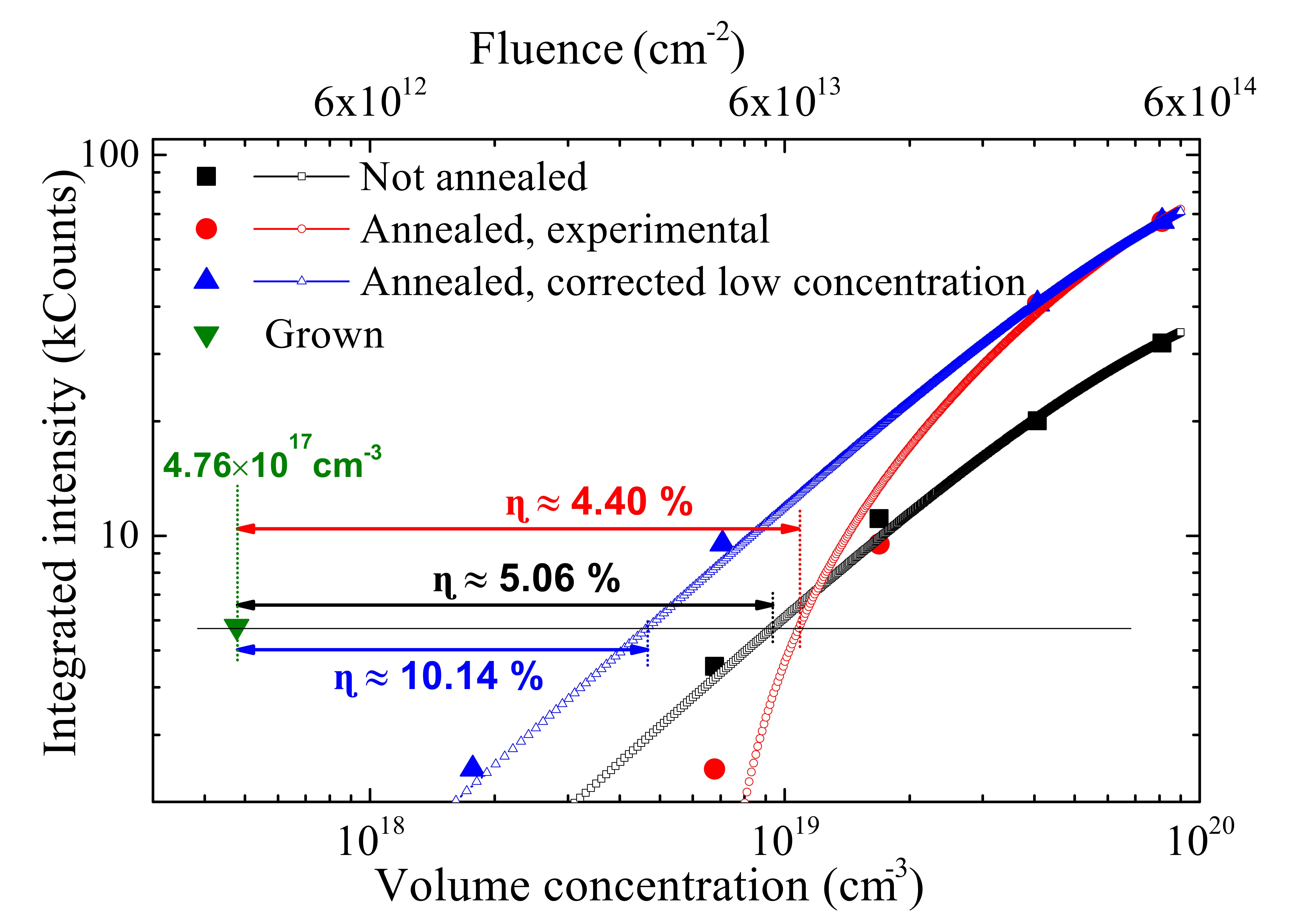}
\caption{(Color online) Fit of the intensity-on-concentration dependence. Experimental data of the implanted sample are related to the doped-as-grown. Parameters of the fitting function are given in Table~\ref{table:tab3}. Implantation efficiency is estimated in the range of 5~\% to 10~\% depending on fluence.}
\label{fig:rubionfit}
\end{figure}

The fits of the dependance of intensity on the fluence are given in Fig.~\ref{fig:rubionfit}. Data points and line given with squares correspond to the non-annealed sample, those marked with circles represent the annealed sample. Upside-down triangle stands for the doped as-grown sample luminescence, corresponding to the luminescent volume equal to the volume occupied by the implanted ions. We take the grown sample luminescence as given by a defined number of ions. Putting this intensity value into the fit-function with the obtained parameters, we find the implanted concentration as $9.45 \times 10^{18}~cm^{-3}$ for the non-annealed sample and as $1.09 \times 10^{19}~cm^{-3}$ for the annealed sample, which gives the same luminescence intensity as the doped as-grown sample. We assume that the doped as-grown sample concentration, $4.76 \times 10^{17}~cm^{-3}$, is the correct one, and we may conclude that $5.06~\%$ ($= \frac{4.76 \times 10^{17}~cm^{-3}}{9.45 \times 10^{18}~cm^{-3}}$) of the implanted ions are optically detectable just after the implantation, where as many as $4.40~\%$ ($= \frac{4.76 \times 10^{17}~cm^{-3}}{1.09 \times 10^{19}~cm^{-3}}$) of the implanted ions turn to be successfully embedded into the crystal lattice.

However, it is quite evident that the activation rate reaches a more constant value of about $2.09$ with higher fluence - which is different from the one taken into our calculation. The obtained efficiency of $4.40~\%$ is no longer a valid number in that range. Therefore, we consider for all performed fluences, the annealing activation had a value of $2.09$. This consideration is represented in Fig.~\ref{fig:rubionfit} with the line marked with triangles. We repeat the calculation of the concentration, that corresponds to the intensity of the grown sample, and obtain $4.71 \times 10^{18}~cm^{-3}$. The implantation efficiency reaches $10.14~\%$ in this case. 

In conclusion, the FIB technique is a suitable technique to locally dope Y$_2$SiO$_5$ crystals with the erbium ions. Luminescence from the implanted ions is detected even before the proper annealing and activation are performed. However, the annealing  process plays a crucial role to properly introduce the ions into the lattice. When high enough damage is created during the implantation, which is true for the fluences above $2 \times 10^{14}~cm^{-2}$, the activation rate of the implanted ions reaches a constant value. The process efficiency might yet be improved by introducing implantation into a hot substrate. 
Thus, this work shows the promising potential of the FIB implantation technique for future application in quantum information networks.

\begin{acknowledgments}
We would like to thank RUBION, the group of Prof. J. Meier, for the given possibility of the measurements. This work was financially supported by the BMBF Quantum communication program, BMBF-QUIMP 16BQ1062. We would like also acknowledge the DFH/UFA CDFA-05-06 Nice-Bochum and RUB Research School for the communication platform provided.
S. P. acknowledges financial support by the LGF of Baden-W\"{u}rttemberg.
\end{acknowledgments}

\bibliographystyle{apsrev4-1}
\bibliography{implanted_optics}

\end{document}